\documentclass[aps,prd,showkeys,superscriptaddress,twocolumn,nofootinbib,floatfix]{revtex4-1}

\usepackage{graphicx,color}
\usepackage{relsize}
\usepackage{slashed}
\usepackage{color}
\usepackage{ulem}
\usepackage{tabu}
\usepackage{bm}

\usepackage{amsmath}
\usepackage{amssymb}
\usepackage{amsthm}
\usepackage{mathrsfs}
\usepackage{graphicx}
\usepackage{epstopdf}
\usepackage{fancyhdr}
\usepackage{array}
\usepackage[all]{xy}
\usepackage{eufrak}
\usepackage{euscript}
\usepackage{enumerate}
\usepackage{slashed}
\usepackage{hyperref}
\usepackage{subfigure} 
\usepackage{epstopdf} 

\hypersetup{pdftex,colorlinks=true,linkcolor=blue,citecolor=blue,menucolor=black,urlcolor=blue,filecolor=blue}

\begin{document}

\title{$ \boldsymbol{\Upsilon(nl)}$ decay into $ \boldsymbol{B^{(*)} \bar B^{(*)}}$ }

\author{Wei-Hong Liang}
\email{liangwh@gxnu.edu.cn}
\affiliation{Department of Physics, Guangxi Normal University, Guilin 541004, China}
\affiliation{Guangxi Key Laboratory of Nuclear Physics and Technology, Guangxi Normal University, Guilin 541004, China}

\author{Natsumi Ikeno}
\email{ikeno@tottori-u.ac.jp}
\affiliation{Department of Life and Environmental Agricultural Sciences, Tottori University, Tottori 680-8551, Japan}
\affiliation{Departamento de F\'{i}sica Te\'{o}rica and IFIC, Centro Mixto Universidad de Valencia - CSIC,
Institutos de Investigaci\'{o}n de Paterna, Aptdo. 22085, 46071 Valencia, Spain.}

\author{Eulogio Oset}
\email{eulogio.oset@ific.uv.es}
\affiliation{Department of Physics, Guangxi Normal University, Guilin 541004, China}
\affiliation{Departamento de F\'{i}sica Te\'{o}rica and IFIC, Centro Mixto Universidad de Valencia - CSIC,
Institutos de Investigaci\'{o}n de Paterna, Aptdo. 22085, 46071 Valencia, Spain.}

\date{\today}

\begin{abstract}
  We have evaluated the decay modes of the $\Upsilon(4s), \Upsilon(3d), \Upsilon(5s), \Upsilon(6s)$ states
  into $B\bar B, B\bar B^*+c.c., B^* \bar B^*, B_s \bar B_s, B_s \bar B^*_s +c.c., B^*_s \bar B_s^* $ using the $^3P_0$ model
  to hadronize the $b\bar b$ vector seed,
  fitting some parameters to the data.
  We observe that the $\Upsilon(4s)$ state has an abnormally large amount of meson-meson components in the wave function,
  while the other states are largely $b\bar b$.
  We predict branching ratios for the different decay channels which can be contrasted with experiment for the case of the $\Upsilon(5s)$ state.
  While globally the agreement is fair,
  we call the attention to some disagreement that could be a warning for the existence of more elaborate components in the state.
\end{abstract}



\maketitle

\section{Introduction}
\label{sec:intro}

The vector $\Upsilon(nl)$ states are a good example to test quark models with $b\bar b$.
The large mass of the $b$ quark makes them excellent nonrelativistic systems
and the theoretical predictions~\cite{Isgur} agree well with experiment,
at least for the first states, with discrepancies in the mass of only a few MeV.
There are many variants of the $b\bar b$ quark model~\cite{Moats,Entem,Santo,Vijande,Gonzalez,vanBev},
producing again similar results for the lowest states and larger discrepancies for the higher excited states.
A more complete view can be obtained from Ref.~\cite{xliu}.
In particular, discussions on non $b\bar b$ configurations for higher excited states are on going.
More concretely, problems stem from the states which can decay into $B^{(*)} \bar B^{(*)}$, starting from the $\Upsilon(4s)$.
In Table \ref{tab:tab1}  we show predictions for masses of these states and current PDG~\cite{pdg} values.
We follow here the PDG assignments, but one should be aware of the different assignments given in some theoretical works~\cite{vanBev,xliu}, particularly the 10579~MeV peak, which is associated in Ref.~\cite{vanBev} to a $B \bar B$ enhancement related to the $\Upsilon(2d)$ state.

\begin{table}[b]
\caption{Predictions of masses for $\Upsilon(nl)$ and PDG values in MeV.}
\centering
\begin{tabular}{c l l l}
\hline\hline
{\bf ~~State} ~~~~& Ref.~\cite{Isgur}~~~~~ & Ref.~\cite{Gonzalez}~~~~~ & PDG\\
\hline
$4s$ ~& $10630$ & $10608$ & $10579$ \\
$3d$ ~& $10700$ & $10682$ & $10753$ \cite{belle}\\
$5s$ ~& $10880$ & $10840$ & $10889$ \\
$4d$ ~&         & $10899$ &     \\
$6s$ ~& $11100$ &         & $10993$ \\
\hline\hline
\end{tabular}
\label{tab:tab1}
\end{table}

The $\Upsilon(3d)$ state, reported recently by the Belle collaboration with mass $10753$ MeV,
is close to the quark model predictions (see Table \ref{tab:tab1}).
Yet, claims that this state could be a tetraquark state are made in Ref.~\cite{gang}.
Similarly, the $\Upsilon(5s)$ state is questioned as a pure $b\bar b$ state in base to the $\pi^+\pi^- \Upsilon(n'l')$ decay rates,
and a mixture of $\Upsilon(5s)$ plus the lowest $1^{--}$ hybrid state \cite{Bruschini} is invoked in Ref.~\cite{pedrohadron}.
The meson-meson components of the charmonium states were studied in detail in Ref.~\cite{Eichten},
using effective interactions for the hadronization.
A similar work is done in Ref.~\cite{vanRupp} for the bottomium states,
addressing the decay modes of the $b\bar b$ states into $B^{(*)} \bar B^{(*)}$.
More recently
there is some work done on the $B^{(*)} \bar B^{(*)} \pi$ decay \cite{SLZhu}.
Closest work would be the ratios predicted for $e^+e^- \to B\bar B$, $e^+e^- \to B\bar B^*+c.c.$ and $e^+e^- \to B^* \bar B^*$ cross sections
using heavy quark spin symmetry in Refs.~\cite{Rujula,Manohar}, but, as mentioned in Ref.~\cite{Voloshin},
these predictions are in conflict with experiment and it was blamed on the proximity of quarkonium resonances  to thresholds of these channels,
suggesting the mixture of the $\Upsilon(nl)$ states with some meson-meson component to solve this conflict \cite{Voloshin}.
In the present work we address these problems for the $4s, 3d, 5s, 6s$ states, for which there are experimental data.

\section{Formalism}
\label{sec:form}

\begin{figure}[b]
\begin{center}
\includegraphics[scale=0.6]{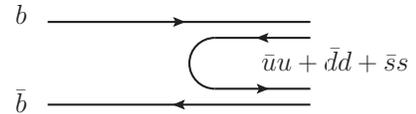}
\end{center}
\vspace{-0.7cm}
\caption{Hadronization of $b\bar b$.}
\label{Fig:1}
\end{figure}

We follow here the formalism used in Refs.~\cite{lineshape,bayar}.
For this we use the $^3P_0$ model to hadronize the $b\bar b$ vector state and generate two $B^{(*)} \bar B^{(*)}$ mesons, as shown in Fig.~\ref{Fig:1},
creating a flavor-scalar state with the quantum numbers of the vacuum.

We consider only $\bar u u +\bar dd+\bar ss$
since the $\bar cc, \bar bb$ components give rise to meson-meson states too far away in energy to be relevant in the process.
If we write the $q \bar q$ matrix, $M$, with the $u, d, s, c$ quarks we have
\begin{equation}\label{eq:1}
M=(q \; \bar q)=\left(
           \begin{array}{cccc}
             u\bar u & u \bar d & u\bar s & u\bar b \\
             d\bar u & d\bar d & d\bar s  & d \bar b \\
             s\bar u & s\bar d & s\bar s  & s \bar b\\
             b\bar u & b\bar d & b\bar s  & b \bar b\\
           \end{array}
         \right),
\end{equation}
and then, after hadronization we find
\begin{equation}\label{eq:bbbar}
  b\bar b \rightarrow \sum_{i=1}^3 \, b \,\bar q_i q_i \,\bar b \, = \sum_{i=1}^3  \, M_{4i}\, M_{i4}.
\end{equation}
If we write $M_{4i}\, M_{i4}$ in terms of the $B, B^*$ mesons, we find the combinations
\begin{align}\label{eq:BBar}
    &B^- B^+ + \bar B^0 B^0 + \bar B_s^0 B_s^0,        \nonumber\\
    &B^- B^{*+} + \bar B^0 B^{*0} + \bar B_s^0 B^{*0}_s, \nonumber\\
    &B^{*-} B^+ + \bar B^{*0} B^0 +\bar B^{*0}_s B^{0}_s, \\
    &B^{*-} B^{*+} + \bar B^{*0} B^{*0} + \bar B_s^{*0} B_s^{*0}.\nonumber
\end{align}
The next step is to see the relationship of the production modes of these four combinations $PP, \, PV, \, VP, \, VV$ ($P$ pseudoscalar, $V$ vector)
and for this we use the $^3P_0$ model \cite{Micu,Oliver}.
The details of the angular momentum algebra involved are shown in Ref.~\cite{lineshape},
and relative weights for $PP, \, PV, \, VP, \, VV$ production are obtained to which we shall come back below.

\begin{figure}[t]
\begin{center}
\includegraphics[scale=0.55]{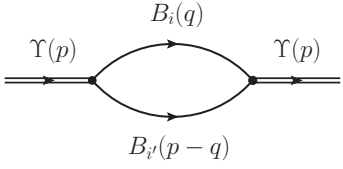}
\end{center}
\vspace{-0.7cm}
\caption{Selfenergy diagram of the $\Upsilon$ accounting for $ii' \equiv B^{(*)} \bar B^{(*)}$ intermediate states.}
\label{Fig:2}
\end{figure}

The formalism that we follow relies on the use of the vector propagator which is dressed including the selfenergy due to $B^{(*)} \bar B^{(*)}$ production,
as shown in Fig.~\ref{Fig:2}.

The renormalized vector meson propagator is written as
\begin{equation}\label{eq:DR}
  D_R= \frac{1}{p^2-M^2_R-\Pi(p)},
\end{equation}
where the selfenergy $\Pi(p)$ is given by
\begin{eqnarray}\label{eq:Pi}
  -i\Pi (p) &=& \int \frac{{\rm d}^4q}{(2\pi)^4}\, (-i) V_1 \, (-i) V_2 \; \frac{i}{q^2-m^2_{B_i}+ i \epsilon} \nonumber\\
   &&\times \frac{i}{(p-q)^2-m^2_{B_{i'}}+ i \epsilon}\; F^2(q),
\end{eqnarray}
and the vertex for $\Upsilon B_i B_{i'}$ is of the type
\begin{equation}\label{eq:vertex}
  V_{R, B_i B_{i'}} = A\, g_{R_{i}}\, |\vec q\,|,
\end{equation}
where $A$ is an arbitrary constant to be fitted to the data
and $g_{R_{i}}$ are weights for the different $B_i B_{i'}$ states which are evaluated using the $^3P_0$ model in Ref.~\cite{lineshape}.
Eq.~\eqref{eq:vertex} is an effective vertex which takes into account the sum over polarizations of the vectors in the $\Pi$ loop in Fig.~\ref{Fig:2} .
For a given channel $B_i B_{i'}$ the selfenergy is then given by
\begin{eqnarray}\label{eq:Pi2}
  \Pi_i (p) &=& i\, g^2_{R_{i}}\, A^2 \int \frac{{\rm d}^4q}{(2\pi)^4}\, \vec q\,^2 \; \frac{1}{q^2-m^2_{B_i}+ i \epsilon} \nonumber\\
   &&~~~~~~~~~~~ \times \frac{1}{(p-q)^2-m^2_{B_{i'}}+ i \epsilon} \; F^2(q),
\end{eqnarray}
where the coefficients $g_{R_{i}}$ are evaluated with the $^3P_0$ model in Ref.~\cite{lineshape}.
The $q^0$ integration is done analytically and we find
\begin{equation}\label{eq:Pi3}
  \Pi (p^0)= A^2 \sum_i g^2_{R_i}\, \tilde{G}_i(p^0),
\end{equation}
where
\begin{eqnarray}
 \tilde{G}_i(p^0) &=& \int  \frac{{\rm d} q}{(2\pi)^2}\; \frac{w_1(\vec q\,)+w_2(\vec q\,)}{w_1(\vec q\,)\, w_2(\vec q\,)} \nonumber\\
   && \times \frac{\vec q\,^4}{(p^0)^2-[w_1(\vec q\,)+w_2(\vec q\,)]^2 +i\epsilon}\, F^2( q),
\end{eqnarray}
where $w_i (\vec q\,)=\sqrt{m_i^2 +\vec q\, ^2}$
and $F(\vec q\,)$ is a form factor that, inspired upon the Blatt-Weisskopf barrier penetration factor \cite{blatt},
we take of the type of Ref.~\cite{lineshape} ($p^0 =\sqrt{s}$),
\begin{align}\label{eq:ff1}
F^2(q)=\frac{1+(R\,q_{\rm on})^2}{1+(R\,q)^2},~~~~q_{\rm on}=\frac{\lambda^{1/2}(s,m_1^2,m_2^2)}{2\sqrt{s}},
\end{align}
with $q_{\rm on}$ taken zero below threshold, and $R$ a parameter to be fitted to data.

In order to have a pole at $M_R$, we define
\begin{align}\label{eq:Piprime}
\Pi^\prime(p)=\Pi(p)-{\rm Re}\Pi(M_R),
\end{align}
which renders $\Pi^\prime(p)$ convergent ($\Pi_i (p)$ of Eq.~\eqref{eq:Pi2} is logarithmically divergent) and then we have the $\Upsilon$ propagator as
\begin{align}\label{eq:DR2}
D_R(p)=\frac{1}{p^2-M^2_R-\Pi^\prime(p)}.
\end{align}

According to Refs.~\cite{Giacosa,lineshape} the cross section for $e^+ e^- \to R\to \sum_i B_i B_{i'}$ is given by
\begin{equation}\label{eq:sigma}
  \sigma =- f^2_R \; {\rm Im} D_R(p),
\end{equation}
and the individual cross section to each channel by
\begin{align}\label{eq:sigmai}
\sigma_i =- f^2_R \;\frac{{\rm Im}\Pi_i(p)}{[p^2-M_R^2-{\rm Re}\Pi'(p)]^2+[{\rm Im}\Pi(p)]^2}.
\end{align}

It is customary to make an expansion of $\Pi'$ in Eq.~\eqref{eq:DR2} around the resonance mass as
\begin{align}\label{g1}
D_R(p)&\!\simeq \!\frac{1}{p^2\!-\!M^2_R\!-\!(p^2-M^2_R)\frac{\partial {\rm Re}\Pi'(p)}{\partial p^2}\big|_{p^2=M^2_R}\!-\!i {\rm Im}\Pi(p)}\nonumber\\
&=\frac{Z}{p^2-M^2_R-iZ\,{\rm Im} \Pi(p)},
\end{align}
with
\begin{align}\label{eq:Z}
Z=\frac{1}{1-\frac{\partial \,{\rm Re}\,\Pi'(p)}{\partial p^2}\big|_{p^2=M^2_R}},
\end{align}
and then the width of the resonance is given by
\begin{equation}\label{eq:Gamma}
  \Gamma = -\frac{1}{M_R}\; Z\, {\rm Im} \Pi(p)\big|_{p^2=M^2_R},
\end{equation}
and for each individual channel
\begin{equation}\label{eq:Gammai}
  \Gamma_i =- \frac{1}{M_R}\; Z\, {\rm Im} \Pi_i(p)\big|_{p^2=M^2_R}.
\end{equation}
The value of $Z$ provides the strength of the $\Upsilon$ vector component (not to be associated to a probability when there are open channels \cite{lineshape,Aceti}).
We shall see that for the $\Upsilon(4s)$ state the $Z$ value is relatively different from $1$,
indicating the importance of the weight of the $B_i B_{i'}$ channels in the state.
If $Z$ is close to $1$ we can make a series expansion of $Z$ in Eq.~\eqref{eq:Z}
\begin{align}\label{eq:Z2}
Z\simeq 1+\frac{\partial\, {\rm Re}\Pi'(p)}{\partial p^2}\Big|_{p^2=M^2_R}= 1+\sum_i \frac{\partial\, {\rm Re}\Pi_i'(p)}{\partial p^2}\Big|_{p^2=M^2_R},
\end{align}
such that each individual value
\begin{equation}\label{eq:Pi}
  P_i =-\frac{\partial\, {\rm Re}\Pi_i'(p)}{\partial p^2}\Big|_{p^2=M^2_R}
\end{equation}
can be interpreted as the weight of each meson-meson component in the $\Upsilon$ wave function
(see Ref.~\cite{Aceti} for the precise interpretation of this quantity,
that gives an idea of the weight of each component but cannot be identified with a probability.).

The values of the couplings $g_{Ri}$,
obtained from the $^3P_0$ model in Refs.~\cite{lineshape,bayar},
are shown in Table \ref{tab:weights}.

\begin{table}[b]
\renewcommand\arraystretch{1.2}
\centering
\caption{\vadjust{\vspace{-2pt}}$g^2_{R_i}$ Weights for the different $B^{(*)} \bar B^{(*)}$ components.}\label{tab:weights}
\begin{tabular*}{0.4\textwidth}{@{\extracolsep{\fill}}lll}
\hline
\hline
     channels                    & $s$-wave  &    $d$-wave  \\
     \hline
    $B^0\bar B^0$                &$1/12$ &  $1/12$         \\
    $B^+B^-$                      &$1/12$ &  $1/12 $ \\
    $B^0\bar B^{*0}$          &$1/6 $  & $1/24 $        \\
    $B^{*0}\bar B^{0}$          &$1/6 $  & $1/24 $        \\
    $B^+B^{*-}$               &$1/6$  &  $1/24$         \\
   $B^{*+}B^{-}$               &$1/6$  &  $1/24$         \\
    $B^{*0}\bar B^{*0}$        &$7/12$  &  $77/120$~~     \\
    $B^{*+}B^{*-}$             &$7/12$  &  $77/120$     \\
    $B_s^0 \bar B^0_s$               &$1/12$  &  $1/12$     \\
    $B_s^0 \bar B_s^{*0}$   &$1/6$  &  $1/24$    \\
   $B_s^{*0} \bar B_s^{0}$   &$1/6$  &  $1/24$    \\
    $B^{*0}_s \bar B^{*0}_s$   &$7/12$  & $77/120$ \\
\hline
\hline
\end{tabular*}
\end{table}

The weights $\frac{2}{12} : \frac{8}{12} : \frac{14}{12} : \frac{1}{12} : \frac{4}{12} : \frac{7}{12}$ for $B\bar B : B\bar B^*+c.c. : B^* \bar B^* : B_s \bar B_s : B^*_s \bar B_s+c.c. : B^*_s \bar B^*_s $ of Table \ref{tab:weights} for the $s$-wave states are in agreement with those quoted in Ref.~\cite{vanBev} (see also formulae in Ref.~\cite{vanRupp}).
The first three ratios are also given in Ref.~\cite{RujulaRatio} in base to heavy quark symmetry.
For $d$-wave we made some approximations in Ref.~\cite{lineshape}.
Indeed, in Eq.~(A8) of Ref.~\cite{lineshape} the possible values of an internal angular momentum are $l=1,3$,
but assuming dominance of the lowest angular momentum,
since the radial matrix elements involve $j_l(qr)$,
$l=3$ is neglected.
Also the coupling involves $Y_{3\mu}(\hat{q})$ rather than $Y_{1\mu}(\hat{q})$ as implicit in Eq.~(\ref{eq:vertex}).
In this case we have some differences with respect to Refs.~\cite{vanBev,vanRupp}.

\section{Results}
\label{sec:res}
The strategy is to fit the parameters $A, R$ to the shape of the $e^+e^- \to \Upsilon(4s) \to B\bar B$ cross section, Eq.~\eqref{eq:sigmai}.
The parameter $f_R$ is irrelevant for the shape.
We also fine tune the value of $M_R$ around the nominal value of the PDG.
Then we take the same value of $R$, which gives a range of the momentum distribution of the internal meson-meson components,
and the parameter $A$ will be adjusted to the width of each of the other states.

\subsection{$\boldsymbol{\Upsilon(4s)}$ state}
\label{subsec:resUpsilon4s}
\begin{figure}[t]
\begin{center}
\includegraphics[scale=0.7]{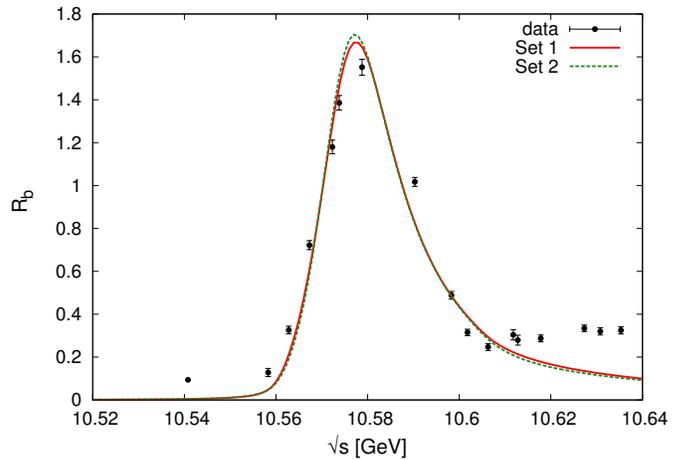}
\end{center}
\vspace{-0.7cm}
\caption{The $e^+e^- \to \Upsilon(4s) \to B\bar B$ cross section and our fits to the data from BaBar \cite{Aubert}. $R_b = 3 s \sigma / 4 \pi \alpha^2$.}
\label{Fig:3}
\end{figure}

In Fig.~\ref{Fig:3}, we show the data of Ref.~\cite{Aubert} for $e^+e^- \to \Upsilon(4s) \to B\bar B$.
The data are for $R_b = \frac{3 s}{4 \pi \alpha^2}\sigma$.
We observe that it is possible to find good fits to the data with different sets of parameters.
We show the results with two significative ones that lead to the parameters
\begin{flalign} \label{eq:parameter1} \text{Set 1: }
  f_R    &=1.05\times 10^{-3},     &   M_R &= 10579\, {\rm MeV},   \nonumber\\
  A        &= 118, &   R     &= 0.007\, {\rm MeV}^{-1};
\end{flalign}
\begin{flalign} \label{eq:parameter2} \text{Set 2: }
  f_R    &=1.3\times 10^{-3},     &   M_R &= 10579\, {\rm MeV},   \nonumber\\
  A        &= 141.4, &   R     &= 0.005\, {\rm MeV}^{-1}.
\end{flalign}

At this point it is relevant to make a comment concerning the values obtained for $R$.
We can intuitively think that these values would give us an idea of the size of the components.
They correspond to about $1\,{\rm fm}$ and $1.38\,{\rm fm}$, for set~1 and set~2,
which we could consider too large.
Actually, this should be attributed to the $B^{(*)} \bar B^{(*)}$ cloud,
not to the seed of the $b\bar b$ quarks.
Then two comments are in order:
the first one is that when dealing with hadron components of composite states,
it was found in Ref.~\cite{qixinchi} that the sizes are bigger than expected,
and components around $1\,{\rm fm}$ are not negligible.
The other comment is that one should not be so strict in associating $R$ to a size.
Eq.~\eqref{eq:ff1} has been used as a regulator in the integral of Eq.~\eqref{eq:Pi2},
which reduces the integrand in two powers of $q$.
Yet, we need an extra regulator,
which is the substraction of Eq.~\eqref{eq:Piprime},
to finally render the selfenergy finite.

In Table~\ref{tab:2} we show the values of $-\frac{\partial \, \Pi_i}{\partial p^2}\Big|_{p^2=M^2_R}$ for each channel and the value of $Z$.
\begin{table}[t]
\renewcommand\arraystretch{1.2}
\centering
\caption{\vadjust{\vspace{-2pt}}Values of $-\frac{\partial \, \Pi_i}{\partial p^2}\Big|_{p^2=M^2_R}$ for the different channels and the value of $Z$ for $\Upsilon(4s)$ state.}\label{tab:2}
\begin{tabular*}{0.46\textwidth}{@{\extracolsep{\fill}}lll}
\hline
\hline
                    & Set 1  &    Set 2 \\
                               \hline
    $B^0\bar B^0$                    &$-0.021 + 0.232i $  &  $0.008 + 0.386i $         \\
    $B^+B^-$                         &$-0.024+0.234i$  &  $0.004 + 0.389i $        \\
    $B^0\bar B^{\ast0}+c.c.$          &$0.080+0.002i$  &   $0.208 + 0.005i $       \\
    $B^+B^{\ast-}+c.c.$               &$0.080+0.002i$  &   $0.209 + 0.005i $         \\
    $B^{\ast0}\bar B^{\ast0}$        &$0.069+0.001i$  &    $0.185 + 0.002i $     \\
    $B^{\ast+}B^{\ast-}$             &$0.069+0.001i$  &  $0.185 + 0.002i $     \\
    $B_s^0 \bar B^0_s$               &$0.005$  & $0.014$       \\
    $B_s^0 \bar B_s^{\ast0}+c.c.$     &$0.015$  &   $0.041$    \\
    $B^{\ast}_s \bar B^{\ast}_s$     &$0.021$  & $0.057$   \\
    Total                            &$0.295+0.472i$  &  $0.912+0.788i$          \\
\hline
    ~~$Z$~~                          & $0.772$      &   $0.523$           \\
\hline
\hline
\end{tabular*}
\end{table}
We can see that the results for the meson-meson weights and the final $Z$ value are different for the two sets, and we must necessarily accept this as uncertainties in our approach. Yet, the message is clear that in that state the strength of the meson-meson components is very large.
The width of the state is $\Gamma =20.5\pm 2.5\, {\rm MeV}$ \cite{pdg},
quite large compared to that of the other $\Upsilon$ states in spite of the limited phase space for the only open channel $B\bar B$.
This feature is what demands a large value of $A$ that translates into a large fraction of the meson-meson components in the $\Upsilon$ wave function.
From Fig.~\ref{Fig:3}, we obtain $\Gamma \sim 20$ MeV, similar to the value quoted in the PDG \cite{pdg},
and a value around $22$ MeV using Eq.~\eqref{eq:Gamma}.
More important than these numbers is that we fit the $B\bar B$ data from BaBar \cite{Aubert}.
We can use Eqs.~\eqref{eq:Gamma} and \eqref{eq:Gammai} to get branching ratios and we obtain the results shown in Table \ref{tab:4sBR}.
\begin{table}[t]
\renewcommand\arraystretch{1.2}
\centering
\caption{\vadjust{\vspace{-2pt}}Branching ratios for $\Upsilon(4s)$ decay.}\label{tab:4sBR}
\begin{tabular*}{0.4\textwidth}{@{\extracolsep{\fill}}lll}
\hline
\hline
    Channel        &    BR$|_{\rm Theo.}$    &   BR$|_{\rm Exp.}$     \\
    $B^0 \bar B^0$ &    $48.8 \%$                     &   $(48.6 \pm 0.6)\%$   \\
    $B^+ B^-$      &    $51.2 \%$                     &   $(51.4 \pm 0.6)\%$  \\
\hline
\hline
\end{tabular*}
\end{table}
The branching ratios obtained are the same for the two sets of parameters.
The good width is a consequence of the fit, and the branching ratios,
in agreement with experiment,
are a consequence of the different phase space for $ B^+ B^-$ and $B^0 \bar B^0$,
due to their different masses.

\subsection{$\boldsymbol{\Upsilon(3d)}$ state}
\label{subsec:resUpsilon3d}

We take the PDG mass $10752.7$ MeV and fit the width of $\Gamma=35.5$ MeV and obtain the only free parameter
\begin{flalign} 
\text{Sets 1 and 2:  } A=11, 
\nonumber
\end{flalign}
which gives rise to a width of $35.8$~MeV (set 1) and $35.7$~MeV (set 2).
With this input we look at the weights for the different channels and we find the results of Table~\ref{tab:3}.
\begin{table}[b]
\renewcommand\arraystretch{1.2}
\centering
\caption{\vadjust{\vspace{-2pt}}Values of $-\frac{\partial \, \Pi_i}{\partial p^2}\Big|_{p^2=M^2_R}$ for the different channels and the value of $Z$ for $\Upsilon(3d)$ state.}\label{tab:3}
\begin{tabular*}{0.45\textwidth}{@{\extracolsep{\fill}}lll}
\hline
\hline
                    & Set 1 &    Set 2  \\
                               \hline
    $B^0\bar B^0$                    &$-0.005 + 0.005i$  &   $-0.005 + 0.005i$       \\
    $B^+B^-$                         &$-0.005 + 0.005i$  &   $-0.005 + 0.005i$      \\
    $B^0\bar B^{\ast0}+c.c.$          &$-0.003 + 0.004i$  &  $-0.003 + 0.004i$        \\
    $B^+B^{\ast-}+c.c.$               &$-0.003 + 0.004i$  &  $-0.003 + 0.004i$          \\
    $B^{\ast0}\bar B^{\ast0}$        &$-0.017 + 0.028i$  &  $-0.017 + 0.030i$       \\
    $B^{\ast+}B^{\ast-}$             &$-0.017 + 0.028i$  &  $-0.017 + 0.030i$      \\
    $B_s^0 \bar B^0_s$               &$-0.0002 + 0.002i$  &  $-0.0001 + 0.002i$      \\
    $B_s^0 \bar B_s^{\ast0}+c.c.$     &~~$0.0002$  &  ~~$0.0003$   \\
    $B^{\ast}_s \bar B^{\ast}_s$     &~~$0.001$  & ~~$0.001$   \\
    Total                            &$-0.050 + 0.076i$  &  $-0.049 + 0.081i$          \\
\hline
    ~~$Z$~~                          & ~~$1.053$           &  ~~$1.052$         \\
\hline
\hline
\end{tabular*}
\end{table}
Interestingly, we find now that the value of $Z$ is very close to $1$ and the weight of the meson-meson components very small.
It is interesting to note that the results are remarkably similar when using set 1 or set 2 of parameters, which makes the results more solid.
We could be surprised that $Z$ is bigger than $1$ and  the individual
meson-meson weights for the open channels are complex and have negative
real part. This is a consequence of the fact that these weights cannot
be interpreted as probabilities. Indeed, as discussed in Refs.~\cite{lineshape,Aceti}, the
individual meson-meson weights correspond to the integral of the wave
function squared with a certain phase prescription (not modulus
squared), which for the open channels is complex.  Even then, this
magnitude measures the strength of the meson-meson components and the
message from these results is that this strength is small and the
$\Upsilon$ remains largely as the original $b \bar b$ component.

Similarly, in Table~\ref{tab:4} we show the branching ratios obtained for each channel, for which there are no experimental data.
It is remarkable the large strength for $B^* \bar B^*$ production in spite of its smaller phase space relative to $B\bar B$ or $B \bar B^*+c.c.$.
We also note that the results with set~1 and set~2 are practically identical for the two sets of parameters of Eqs.~(\ref{eq:parameter1}), (\ref{eq:parameter2}).
This is the only case of a $d$-state that we analyze. We already mentioned that for this case we neglect an $l=3$ contribution.
Although we gave qualitative arguments on why we think this contribution is relatively small,
we should be aware of some extra uncertainties compared to the $s$-wave cases.
Further measurements of the branching ratios deduced in Table~\ref{tab:4} will further clarity this issue.
\begin{table}[t]
\renewcommand\arraystretch{1.2}
\centering
\caption{\vadjust{\vspace{-2pt}}Branching ratios of $\Upsilon(3d)$ decaying to different channels.}\label{tab:4}
\begin{tabular*}{0.25\textwidth}{@{\extracolsep{\fill}}ll}
\hline
\hline
    Channel                  &    BR$|_{\rm Theo.}$        \\
\hline
    $B\bar B$                &    $21.3 \%$                \\
    $B \bar B^*+c.c.$        &    $14.3 \%$                 \\
    $B^* \bar B^*$           &    $64.1 \%$                \\
    $B_s \bar B_s$           &    $0.3 \%$                \\
\hline
\hline
\end{tabular*}
\end{table}

\subsection{$\boldsymbol{\Upsilon(5s)}$ state}
\label{subsec:resUpsilon5s}
We take the nominal PDG mass $M_R=10889.9$ MeV and fit $A$ to get the width of $41.4\; {\rm MeV}$, which is the $81.25\%$ of the $\Upsilon(5s)$ width $51$ MeV.
The value $81.25\%$ is the sum of the experimental branching ratios of $\Upsilon(5s)$ decaying to the different $B^{(*)} \bar B^{(*)}$ channels.
We obtain
\begin{equation*}
 \text{Sets 1 and 2:  } A = 5.64.
\end{equation*}
The weights of the meson-meson components and the value of $Z$ are shown in Table~\ref{tab:5}.
\begin{table}[b]
\renewcommand\arraystretch{1.2}
\centering
\caption{\vadjust{\vspace{-2pt}}Values of $-\frac{\partial \, \Pi_i}{\partial p^2}\Big|_{p^2=M^2_R}$ for the different channels and the value of $Z$ for $\Upsilon(5s)$ state.}\label{tab:5}
\begin{tabular*}{0.35\textwidth}{@{\extracolsep{\fill}}lcc}
\hline
\hline
                    & Set 1 &    Set 2  \\
                               \hline
    $B^0\bar B^0$                    &$-0.003 + 0.002i$  &    same      \\
    $B^+B^-$                         &$-0.003 + 0.002i$  &         \\
    $B^0\bar B^{\ast0}+c.c.$          &$-0.009 + 0.006i$  &          \\
    $B^+B^{\ast-}+c.c.$               &$-0.009 + 0.006i$  &            \\
    $B^{\ast0}\bar B^{\ast0}$        &$-0.012 + 0.010i$  &       \\
    $B^{\ast+}B^{\ast-}$             &$-0.012 + 0.010i$  &      \\
    $B_s^0 \bar B^0_s$               &$-0.001 + 0.001i$  &       \\
    $B_s^0 \bar B_s^{\ast0}+c.c.$     &$-0.003 + 0.004i$  &     \\
    $B^{\ast}_s \bar B^{\ast}_s$     &$-0.002 + 0.005i$  &    \\
    Total                            &$-0.053 + 0.044i$  &        \\
\hline
    ~~$Z$~~                          & $1.056$           &        \\
\hline
\hline
\end{tabular*}
\end{table}
Once again we find that $Z$ is very close to $1$ and the weights of the meson-meson components are very small. Sets 1 and 2 give rise to indistinguishable results.
In Table~\ref{tab:6} we show the branching ratios that we obtain for the different channels and in this case we can compare with the experimental values.
\begin{table}[t]
\renewcommand\arraystretch{1.2}
\centering
\caption{\vadjust{\vspace{-2pt}}Branching ratios for different channels for $\Upsilon(5s)$.}\label{tab:6}
\begin{tabular*}{0.4\textwidth}{@{\extracolsep{\fill}}lll}
\hline
\hline
    Channel                  &    BR$|_{\rm Theo.}$    &   BR$|_{\rm Exp.}$     \\
    $B\bar B$                &    $8.6 \%$            &      $(5.5 \pm 1)\%$\\
    $B \bar B^*+c.c.$        &    $27.8 \%$            &    $(13.7\pm 1.6) \%$ \\
    $B^* \bar B^*$           &    $37.8 \%$            &     $(38.1\pm 3.4) \%$\\
    $B_s \bar B_s$           &    $1.4 \%$            &     $(5\pm 5)\times 10^{-3}$\\
    $B_s \bar B_s^*+c.c.$    &    $ 3.3 \%$            &     $(1.35\pm 0.32)\%$\\
    $B_s^* \bar B_s^*$       &    $2.3 \%$            &    $(17.6\pm 2.7) \%$  \\
\hline
    Total                    &     $81.2 \%$                   & $81.25 \%$ \\
\hline
\hline
\end{tabular*}
\end{table}
Once again there is no difference using set 1 and set 2.
Globally, the branching ratios obtained agree in a fair way with experiment.
We confirm the small $B\bar B$ branching fraction, in spite of the largest phase space,
and the dominance of the $B^*\bar B^*$ channel.
We also find the branching rations for $B_s \bar B_s$ and $B_s \bar B_s^*+c.c.$ channels small like in the experiment.
The only large discrepancy is in the $B^*_s \bar B^*_s$ channel,
where our results are notably smaller than the experimental value.
This large experimental result is not easy to understand.
By analogy to the experimental values for $B\bar B^*+c.c.$ and $B^* \bar B^*$,
using the ratio of these branching ratios ($38.1/13.2$) and multiplying  this value by the experimental branching ratio of $B_s \bar B_s^*+c.c.$,
we should expect a value for the branching ratio of $B^*_s \bar B^*_s$ smaller than $4\%$ because of the reduced phase space.

\subsection{$\boldsymbol{\Upsilon(6s)}$ state}
\label{subsec:resUpsilon6s}
We take again the PDG mass of $M_R=10992.9 ^{+10.0}_{-3.1}$ MeV and the width of $49^{+9}_{-15}$ MeV and make a fit to the width,
assuming that all of it comes from the $B_i \bar B_{i'}$ decay channels (there is no information on these decay channels in the PDG).
We obtain
\begin{equation*}
 \text{Sets 1 and 2:  } A = 4.58.
\end{equation*}
The weights of the different components and the value of $Z$ are shown in Table~\ref{tab:7}.
Once again we see that the value of $Z$ is very close to $1$ and the weight of the meson-meson components very small. Sets 1 and 2 give the same results.
In Table~\ref{tab:8} we show the branching ratios assuming that the width is exhausted by the $B_i B_{i'}$ channels. Again there is no difference between sets 1 and 2.
\begin{table}[t]
\renewcommand\arraystretch{1.2}
\centering
\caption{\vadjust{\vspace{-2pt}}Values of $-\frac{\partial \, \Pi_i}{\partial p^2}\Big|_{p^2=M^2_R}$ for the different channels and the value of $Z$ for $\Upsilon(6s)$ state.}\label{tab:7}
\begin{tabular*}{0.35\textwidth}{@{\extracolsep{\fill}}lcc}
\hline
\hline
                    & Set 1  &    Set 2  \\
                               \hline
    $B^0\bar B^0$                    &$-0.004 + 0.001i$  &    same      \\
    $B^+B^-$                         &$-0.004 + 0.001i$  &         \\
    $B^0\bar B^{\ast0}+c.c.$          &$-0.011 + 0.004i$  &          \\
    $B^+B^{\ast-}+c.c.$               &$-0.011 + 0.004i$  &            \\
    $B^{\ast0}\bar B^{\ast0}$        &$-0.015 + 0.007i$  &       \\
    $B^{\ast+}B^{\ast-}$             &$-0.015 + 0.007i$  &      \\
    $B_s^0 \bar B^0_s$               &$-0.001 + 0.001i$  &       \\
    $B_s^0 \bar B_s^{\ast0}+c.c.$     &$-0.004 + 0.003i$  &     \\
    $B^{\ast}_s \bar B^{\ast}_s$     &$-0.005 + 0.005i$  &    \\
    Total                            &$-0.069 + 0.035i$  &          \\
\hline
    ~~$Z$~~                          & $1.075$           &          \\
\hline
\hline
\end{tabular*}
\end{table}
\begin{table}[t]
\renewcommand\arraystretch{1.2}
\centering
\caption{\vadjust{\vspace{-2pt}}Branching ratios for different channels for $\Upsilon(6s)$.}\label{tab:8}
\begin{tabular*}{0.3\textwidth}{@{\extracolsep{\fill}}ll}
\hline
\hline
    Channel                  &    BR$|_{\rm Theo.}$        \\
\hline
    $B\bar B$                &    $9.0 \%$                \\
    $B \bar B^*+c.c.$        &    $30.7 \%$                 \\
    $B^* \bar B^*$           &    $44.7 \%$                \\
    $B_s \bar B_s$           &    $2.1 \%$                \\
    $B_s \bar B_s^{\ast}+c.c.$ &   $6.2 \%$                \\
    $B^{\ast}_s \bar B^{\ast}_s$  &  $7.3 \%$                \\
\hline
\hline
\end{tabular*}
\end{table}
We should nevertheless mention that the proximity of the $\bar B B_1(5721)(1^+)$ threshold at $11000$ MeV to the mass of this resonance
could have some effect on the selfenergy $\Pi$.
However, if the channel is closed for decay it can affect ${\rm Re} \Pi$
but not the individual ${\rm Im}\Pi_i$ for the open channels needed in the evaluation of the individual decay rates that we have calculated.
This, of course, is the consequence of neglecting the $ B_1(5721) $ width of about $30$ MeV,
and in principle, a favored coupling of $s$-wave to the $1^{--}$ $b\bar b$ state.
So, some uncertainties from this source could be expected.
It will be interesting to compare our results with data when they become available.

\subsection{Discussion}
After we have shown the results, we can say something about the possible mixing of $s$-$d$ components.
Since what matters is the final quantum numbers of the meson, $1^{-\,-}$,
there can be mixing of $b\bar b$ $s$-wave and $b\bar b$ $d$-wave components,
through the intermediate $B^{(*)} \bar B^{(*)}$ states,
which, as we have seen, couple to both configurations.
After the results obtained,
given the small meson-meson components of the $3d, 5s, 6s$,
we can also conclude that the mixing would be small,
and, concerning the $B^{(*)} \bar B^{(*)}$ meson cloud,
these states are largely $b\bar b$ states.
This small mixing has also been reported in previous work on $c\bar c$ states \cite{Eichten,vanZeit}.
We could have an exception with the $4s$ state,
since we showed that the coupling to the $B^{*} \bar B^{*}$ components is so large.
This could give rise to also a relevant mixing with other components.
Actually, the singularity of this state has been widely discussed in some works \cite{vanRupp,vanBev,vanZeit,vrdos}, where the peak at $10580$ MeV is associated to a threshold phenomenon due to the opening of the $B\bar B$ channel and enhanced by the nearby $\Upsilon(2d)$ state, and not to the $4s$ $b\bar b$ state.
The $4s$ state would then be associated to a peak at $10735$ MeV in the $e^+ e^-$ cross section.
There is hence a coincidence with our approach in the fact that the state at $10586$ MeV has a small $b\bar b$ component and the peak contains large meson-meson components.
In our picture, as we see in Table \ref{tab:2},
the largest weights correspond to the closed $B\bar B^* +c.c.$ components.

We should mention that the situation around the peaks and possible states in the region of $10600$-$10700$ MeV is a bit confusing.
Apart from the states reported in Table \ref{tab:tab1},
a state around $10684$ MeV was claimed by CLEO Collaboration \cite{CLEO} and was associated to a $b\bar b g$ state \cite{Ono}.
For our evaluations here we followed the standard PDG classification.

\section{Conclusions}
\label{sec:conc}
We have studied the $B \bar B$, $B \bar B^* +c.c.$, $B^* \bar B^*$, $B_s \bar B_s$, $B_s \bar B_s^*+c.c.$, $B^*_s \bar B_s^*$ decay modes
of the $\Upsilon(4s)$, $\Upsilon(3d)$, $\Upsilon(5s)$, $\Upsilon(6s)$ states of the PDG using the $^3P_0$ model to produce two mesons from the original $b\bar b$ vector state.
We observed interesting things.
The first one is that the $\Upsilon(4s)$ state has an abnormally large width that has as a consequence that
the state contains a relatively large admixture of meson-meson components in its wave function.
It is one exception to the general rule for vector mesons which are largely $q\bar q$ states \cite{Pelaez}.
On the other hand, the other three states studied had very small meson-meson components and the states remain largely in the original $b\bar b$ seed.

We could only test predictions on branching ratios for the $\Upsilon(4s)$ and $\Upsilon(5s)$ states.
In the first case only the $B\bar B$ channel is open and the branching ratios for $B^+ B^-$ and $B^0 \bar B^0$ are determined by phase space in agreement with experiment.
The other case is the $\Upsilon(5s)$ where there are data on the branching ratios.
The agreement is qualitatively good, with the notable exception of the $B^*_s \bar B_s^*$ channel,
and also a factor of two discrepancy in the relative rates of the $B\bar B^*+c.c.$ and $B^*\bar B^*$ channels.
It would be interesting to see if these discrepancies are a warning that more elaborate components for this state, as suggested in Ref.~\cite{pedrohadron}, are at work.

The predictions made for branching ratios for the $\Upsilon(3d)$ and $\Upsilon(6s)$ states should serve as a motivation to measure these magnitudes that will help in advancing our understanding of these bottomium states.

\begin{acknowledgments}
We thank Pedro Gonzalez and E. Eichten for useful discussions.
N. I. acknowledges the hospitality of the Guangxi Normal University,
China, where this work was partly carried out.
This work is partly supported by the National Natural Science Foundation of China under Grants No. 11975083, No. 11947413, 
No. 11847317 and No. 11565007.
The work of N. I. was partly supported by JSPS Overseas Research
Fellowships and JSPS KAKENHI Grant Number JP19K14709.
This work is partly supported by the Spanish Ministerio de
Economia y Competitividad and European FEDER funds under Contracts No. FIS2017-
84038-C2-1-P B and No. FIS2017-84038-C2-2-P B, and the Generalitat Valenciana in the
program Prometeo II-2014/068, and the project Severo Ochoa of IFIC, SEV-2014-0398.
\end{acknowledgments}


  \end{document}